\documentstyle[aps,prb,12pt,amssymb]{revtex}
\begin{document}
\draft
\title{Shot Noise of Single-Electron Tunneling in 1D Arrays}
\author{K. A. Matsuoka and K. K. Likharev}
\address{Department of Physics and Astronomy\\
State University of New York\\
Stony Brook, NY 11794-3800}
\maketitle

\vspace{1em}
\begin{abstract}
We have used numerical modeling and a semi-analytical calculation
method to find the low frequency value $S_{I}(0)$ of the spectral
density of fluctuations of current through 1D arrays of small tunnel
junctions, using the ``orthodox theory'' of single-electron
tunneling. In all three array types studied, at low temperature
\mbox{($k_BT\ll eV$)}, increasing current induces a crossover from the
Schottky value $S_{I}(0)=2e\langle \bar{I}\rangle $ to the ``reduced
Schottky value'' $S_{I}(0)=2e\langle \bar{I}\rangle /N$ (where $N$ is
the array length) at some crossover current $I_{c}$. In uniform arrays
over a ground plane, $I_{c}$ is proportional to $\exp (-\lambda N)$,
where $\lambda ^{-1}$ is the single-electron soliton length. In arrays
without a ground plane, $I_{c}$ decreases slowly with both $N$ and
$\lambda $ . Finally, we have calculated the statistics of $I_{c}$ for
ensembles of arrays with random background charges. The standard
deviation of $I_{c}$ from the ensemble average $\langle I_{c}\rangle $
is quite large, typically between 0.5 and 0.7 of $\langle
I_{c}\rangle$, while the dependence of $\langle I_{c}\rangle $ on $N$
or $\lambda $ is so weak that it is hidden within the random
fluctuations of the crossover current.
\end{abstract}


\newpage

\section{Introduction}

\label{introduction}

Single-electron tunneling (for general reviews, see Refs. %
\onlinecite{mes,sct}) is one of the most active areas of solid state physics
research, but it has several key problems that have not yet been addressed
in detail. One of these problems is the dilemma of the
discreteness/continuity of electric charge transfer\cite{mes}. Several
single-electron tunneling phenomena can be understood as an interplay
between {\it discrete} transfer of electric charge via electron tunneling,
and {\it continuous} transfer in ordinary diffusive conductors. For example,
a tunnel junction biased by a dc current may generate single-electron
tunneling (SET) oscillations with average frequency \cite{av-likh:set}

\begin{equation}  \label{f-set}
f_{S}=\frac{\langle I\rangle }e\,.
\end{equation}

\noindent This effect may be interpreted as a gradual accumulation of
continuous charge on the junction capacitance, followed by the abrupt
passage of one electron through the junction, as soon as the accumulated
charge has reached a threshold level\cite{av-likh:set} $Q_{t}$ $=\pm e/2\,$. 

However, if the charge transfer in the external circuit (fixing the current $%
I$) is discrete, SET oscillations do not exist. This can be seen from the
following general formula for the SET oscillation linewidth \cite
{mes,av-likh:set}: 
\begin{equation}
\Gamma _{S}=\left( \frac{\pi }{e}\right) ^{2}S_{I}(0)\qquad (\Gamma _{S}\ll
f_{S}),
\end{equation}

\noindent where $S_I(f)$ is the spectral density of the bias current
fluctuations. For example, if the current is fixed using another tunnel
junction, then at low temperatures the fluctuations obey the Schottky
formula 
\begin{equation}
S_I(0)=2e\langle I\rangle \,,
\end{equation}

\noindent and $\Gamma _{S}\geq f_{S}$, that is, SET oscillations are
completely smeared by the current fluctuations\cite{mes}. On the other hand,
in macroscopic diffusive conductors, the current noise may be much lower
than the Schottky value, and SET oscillations (and very similar ``Bloch''
oscillations\cite{likh:1985}) may have a relatively narrow bandwidth - see,
e.g., experimental measurements in Ref. \onlinecite{kuzm:1991}.

More generally, discreteness of charge transfer will certainly be one of the
central issues facing the emerging nanoelectronics. This is why it is
important to formulate the conditions under which the transport of charge
through a conductor may be considered as (quasi)continuous, i.e., having 
discreteness $\delta Q\ll e$. The definition of $\delta Q$ can be most
readily introduced in the most interesting case of negligible thermal and
quantum fluctuations: 
\begin{eqnarray}
k_{B}T &\ll &\left[ e\langle V\rangle ,\,eV_{t}\right] \,,  \label{lowT} \\
G &\ll &\frac{e^{2}}{h}\,,  \label{lowG}
\end{eqnarray}
\noindent where $V_{t}$ is the Coulomb blockade threshold of the conductor
and $G$ its effective conductance. However, even in this simplest case, the
definition depends essentially on the characteristic time scale $\tau $ of
an experiment:

A. If $f_{S}\tau \gg 1$ (i.e. either the time scale $\tau $ is large, or the
dc current is high, or both), $\delta Q$ may be defined as follows: 
\begin{equation}
\frac{\delta Q}{e}=\frac{S_{I}(0)}{2e\langle I\rangle }\,.  \label{deltaQ}
\end{equation}
\noindent In fact, if the charge $Q$ transferred through a system may be
presented as a Poissonian series of jumps of fixed height $\delta Q$, then
repeating the well-known derivation of the Schottky formula we arrive at
Eqn. (\ref{deltaQ}). If the jump height is random as well, Eqn. (\ref{deltaQ}%
) is still applicable as an estimate of the average jump height.

B. In the opposite limit, when $\tau $ is much shorter than the average
spacing between the charge jumps ($f_{S}^{-1}$), we are essentially dealing
with the Coulomb blockade regime. In this case an adequate definition of $%
\delta Q$ is as follows: 
\begin{equation}
\delta Q=C_{in}V_{t},
\end{equation}
\noindent where $C_{in}$ is the effective input capacitance of the system of
interest. ($\delta Q$ given by this formula is the fraction of the initial
electric charge of the system which cannot relax, due to the Coulomb
blockade; for a diffusive conductor $\delta Q\to 0$, while for a single
tunnel junction $\delta Q=e/2$.)

One of the most interesting systems capable of quasicontinuous charge
transfer is the 1D array of small tunnel junctions---see, e.g., the review
in Ref. \onlinecite{delsing}. The key property of such an array\cite
{bakh:1989} is that each additional electron inserted into one of its
islands creates a series of gradually decreasing polarization charges, and
hence may be considered as a ``single-electron soliton'' with a
characteristic size $M$ which may be much larger than one island. As a
result, when an electron is drifting along the array, tunneling between the
neighboring islands, the effective charge $Q$ transferred through the
external electrodes---which is essentially what interests us---changes in
jumps of scale $\delta Q\sim e/M$ which may be much smaller than $e$.

This does not mean, however, that the condition $M\gg 1$ is sufficient for
the quasicontinuous transfer of charge in the arrays. In fact, the
calculation of the Coulomb blockade threshold for arrays with uniform\cite
{bakh:1989} and random\cite{middleton,melsen,koro:1996} background charges
have shown that, under definition B, charge transport in uniform arrays with
vanishing background charge {\it cannot} be considered as continuous ($%
\delta Q\sim e$). The reason is that a relatively strong Coulomb blockage
results from the pinning of single-electron solitons by the sharp edges of
the array. Presently, we know only two cases when charge transport through a
uniform array is quasicontinuous ($\delta Q\ll e$) according to the
definition B:

1) An array with arbitrary capacitances, but with the background charge of
all the islands equal to $\pm e/2$. (In this case the Coulomb blockade
threshold vanishes, and $\delta Q\rightarrow 0$.)

2) An array with $1\ll N\ll 2M$ and random background charges. (In this case%
\cite{melsen,koro:1996} $V_{t}\approx 0.5eN^{1/2}/C$, where $C$ is the
tunnel junction capacitance, while $C_{in}=C/N$, and hence $\delta
Q/e\approx 1/2N^{1/2}\ll 1$.)

However, even in these cases, the charge transfer is not automatically
continuous in the sense of criterion A. If the Coulomb blockade is finite,
and the array is driven with dc voltage $V$ slightly above the blockade
threshold $V_{t}$, one of the junctions presents a bottleneck to the
single-electron soliton drift along the array. As a result, the passage of
an electron consists of a long wait at the bottleneck junction, followed by
a rapid burst of transitions through the remaining junctions of the array.
At $V\rightarrow V_{t}$ the statistics of these bursts is always Poissonian,
and the charge transferred by each burst is equal to $e$, so that the shot
noise is well described by the Schottky formula.

Thus, current noise in 1D arrays presents an important problem. To our
knowledge, this problem has previously only been solved\cite{noise} for a
very particular case of uniform arrays with $M\rightarrow \infty$ (zero
stray capacitance) and zero background charge. The objective of this work
was to calculate $S_I(0)$ (and hence the effective discreteness of charge
transfer for long time intervals) for a much broader range of array
parameters.

\section{Basic Formulas}

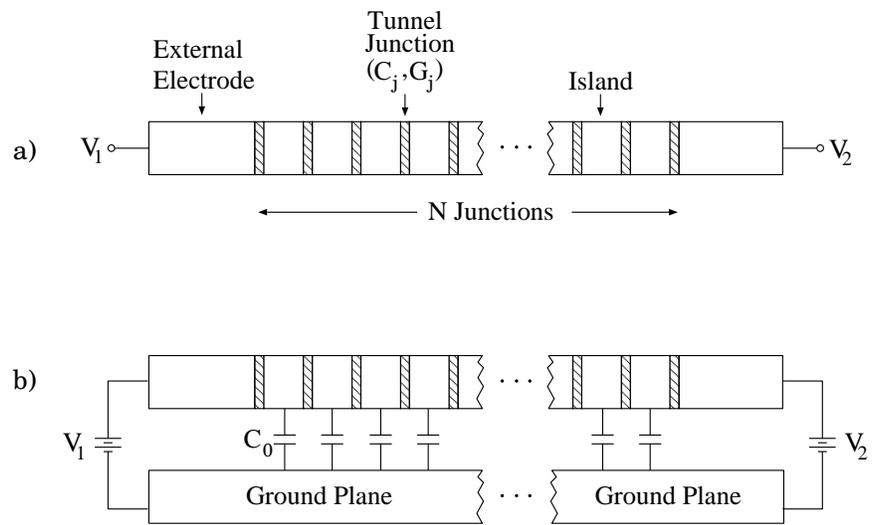
\begin{figure}[tbp]
\caption{1D array of tunnel junctions: (a) general schematic, (b) ``ground
plane'' or ``tridiagonal model'' schematic.}
\label{array-diag}
\end{figure}

We have considered arrays consisting of $N-1$ small metallic islands
connected by $N$ tunnel junctions, and flanked at either end by dc
voltage-biased electrodes (Fig. \ref{array-diag}). Under the conditions
expressed by Eqs. (\ref{lowT}), (\ref{lowG}), we can ignore the effects of
cotunneling and of thermally activated tunneling. Current flow in
single-electron arrays may be analyzed in terms of the junction tunneling
rates, $\Gamma _{ij}$. According to the orthodox theory\cite{bakh:1989}, at
zero temperature,

\begin{equation}  \label{gamma-zero}
\Gamma_{ij} = \left\{ 
\begin{array}{ll}
{\displaystyle \frac{G\Delta W_{ij}}{e^2}} & \Delta W_{ij} > 0 \,, \\ 
0 & \Delta W_{ij} < 0 \,,
\end{array}
\right.
\end{equation}

\noindent where $\Delta W_{ij}$ is the drop in the free (electrostatic)
energy, caused by the tunneling event. The drop in energy due to an electron
tunneling from island $i$ to island $j$ can be written

\begin{equation}
\Delta W_{ij}=e^{2}C_{ij}^{-1}-\frac{e^{2}C_{ii}^{-1}+e^{2}C_{jj}^{-1}}{2}%
+e[\phi_{j}-\phi_{i}]\,,  \label{deltaU}
\end{equation}

\noindent where $\vec{\phi}$ is the vector of the electrostatic potential of
the islands before the jump, and the matrix {\boldmath $C^{-1}$of inverse
capacitances} is defined by the following equation:

\begin{equation}
\phi _{i}=\sum_{j\in isl}C{_{ij}^{-1}(q_{j}+\tilde{q}_{j})}\,,\quad \tilde{q}%
_{j}\equiv \sum_{k\in ext}{\tilde{C}_{jk}V_{k}}\,.
\end{equation}

\noindent Here the matrix {\boldmath $\tilde{C}$} represents
capacitances between islands and external terminals with potentials
$V_{k}.$  If these potentials do not change in time, the probability
that the system preserves its charge state can be expressed
explicitly:

\begin{equation}
P(t)=\exp(-\Gamma(t-t_o)) \,,
\end{equation}

\noindent where $t_{o}$ is the time of the preceding tunneling event, while $%
\Gamma $ is the total rate for all possible tunneling events:

\begin{equation}  \label{gamma-sum}
\Gamma=\sum_{\{ij\} \in jct}\Gamma_{ij}\,.
\end{equation}

\section{Noise Computation}

The preceding relations were incorporated into our main computational tool,
a C{\tt ++} program called {\tt mso}\cite{mso} that uses a Monte Carlo
algorithm\cite{bakh:1989} to simulate the flow of current in dc voltage
biased 1D arrays. The basic unit of calculation in {\tt mso} is the
``current run'', in which charge flows through the array until a
user-specified total charge $Q$ is transferred. To calculate $S_{I}(0)$, 
{\tt mso} loops through a user-specified number $N_{r}$ of current runs,
each starting with the same initial charge state and ending when the total
charge transferred equals $Q$. The same random number generators\cite
{recipes} for time and jump location are used continuously through all
loops. To the extent that the random numbers constitute a ``good''
quasi-random series, the ensemble of current runs represents a statistical
ensemble of independent, initially identical systems.

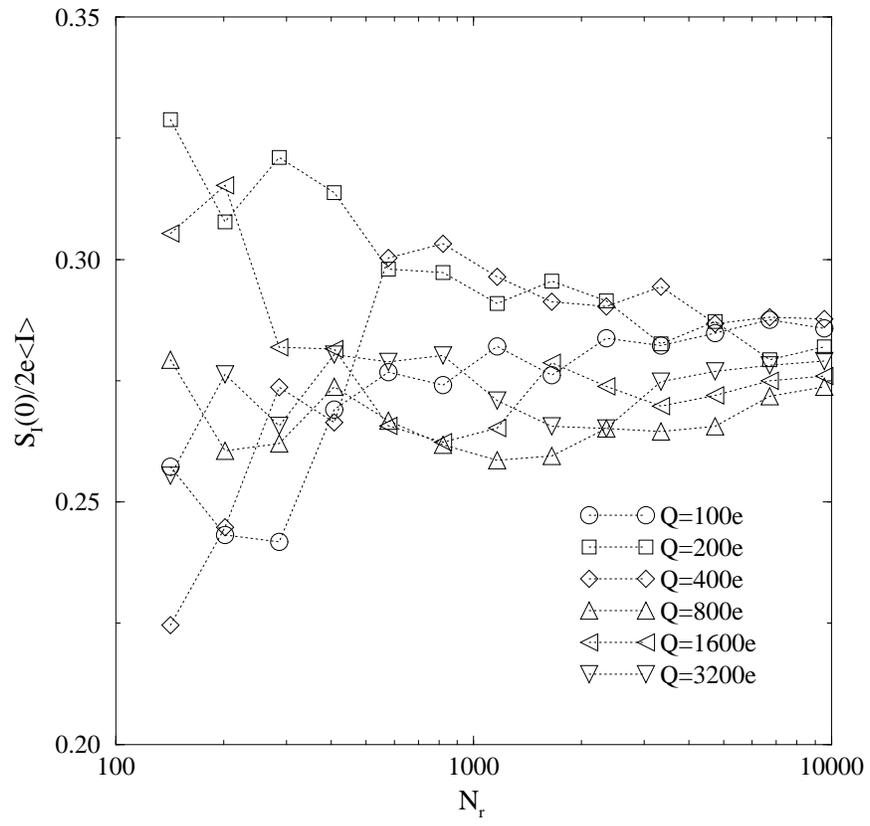
\begin{figure}[tbp]
\caption{Convergence of $S_I(0)/2eI$ calculated by {\tt mso} as a function
of $N_r$, the number of current runs in the statistical ensemble, for
various values of total transferred charge $Q$. Results for an array near
ground plane, $N=20$, $\lambda=0.5$, at crossover current.}
\label{s-qn}
\end{figure}

We may calculate $S_{I}(0)$ from the statistical properties of the
time $T_{Q}$ taken by each run, as\cite{noise}:

\begin{equation}
\frac{S_{I}(0)}{2e\langle \bar{I}\rangle }\approx \frac{Q}{e}\frac{%
\mbox{\rm
Var}(T_{Q})}{\langle T_{Q}\rangle ^{2}}\,,  \label{s-vart}
\end{equation}

\noindent where $\bar{I}=Q/T_{Q}$, and the angle brackets and variance refer
to our statistical ensemble of current runs. Equation (\ref{s-vart}) is
exact only for infinite $Q$ and $N_{r}$; since the jumps are not completely
independent, the accuracy of this formula should be determined
experimentally. Figure~\ref{s-qn} shows a typical dependence of $%
S_{I}(0)/2e\langle \bar{I}\rangle $ on $Q$ and $N_{r}$. The results for $%
Q\gtrsim 1000e$, $N_{r}\gtrsim 1000$ seem to be accurate to within $10\%$ of
the asymptotic value. In this paper, we used the parameters $Q=1000e$, $%
N_{r}=1000$ for calculating shot noise in arrays without background charge,
and assigned $10\%$ error bars to these numbers.

For arrays with random background charge, each point was calculated for 50
different realizations of the background charge for each circuit, using the
parameters $Q=200e$ and $N_{r}=200$, to keep the simulation time within
reasonable limits. Although these calculations are therefore less accurate,
perhaps only to within $20\%$ of their asymptotic value, this inaccuracy was
overshadowed by the overall spread in shot noise values among the different
background charge realizations.

Calculation (CPU) times in {\tt mso} scale as $\sim \!N^{a}QN_{r}$, with $a$
slightly larger than 2. A typical calculation with $N=20$, $Q=1000e$, and $%
N_{r}=1000$ takes around 400 seconds of CPU time on an AlphaStation 250 (266
MHz Alpha, Digital Unix 4.0b, Digital cxx) or around 950 seconds on a Linux
PC (120 MHz Pentium, RedHat Linux 2.0.30, \mbox{Gnu c++}).

\section{Crossover Current}

\label{crossover} 
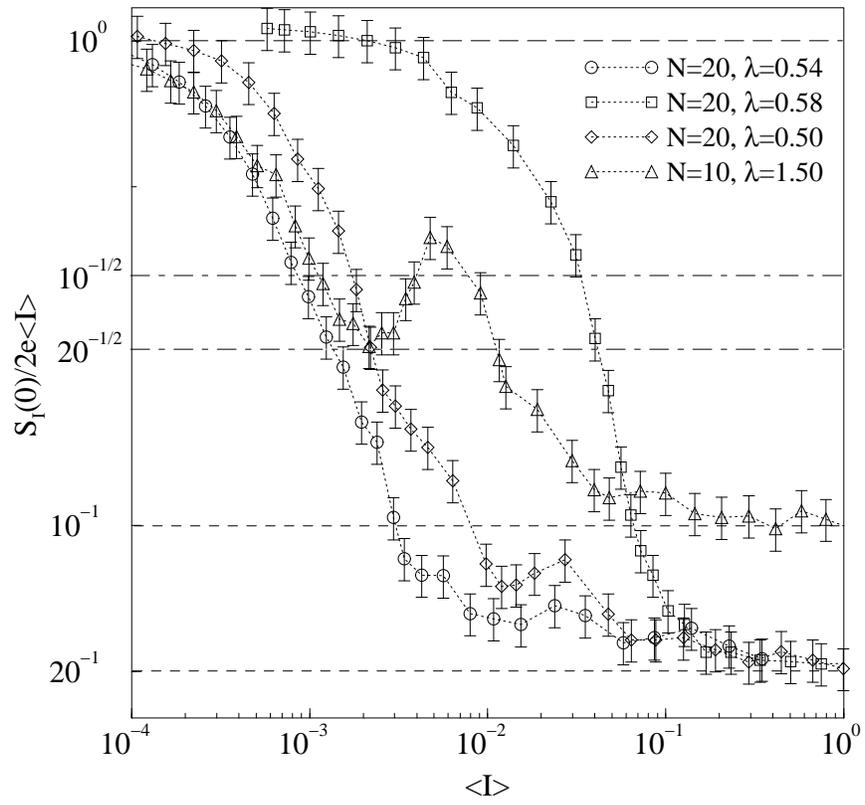
\begin{figure}
\caption{Parametric plots of $S_{I}(0)/2eI$ vs. $I$. Crossover current $I_{c}
$ is defined as the current at which $S_{I}(0)/2eI=1/\protect\sqrt{N}$.
Circles: array near ground plane. Squares: array without ground plane.
Diamonds and triangles: arrays with random background charge. Dotted lines
are only guides for the eye.}
\label{s-i}
\end{figure}

Varying the bias voltage across an array, we have
calculated the average current and spectral density as functions of applied
voltage and have made parametric plots of $S_{I}(0)/2e\langle \bar{I}\rangle 
$ vs. $\langle \bar{I}\rangle$ (Fig. \ref{s-i}). We will refer to these
plots as $S-I$ curves.

The most immediate, universal result of our calculations is the crossover of 
$S_{I}(0)/2e\langle \bar{I}\rangle $ from $1$ to $1/N$ with increasing
current. This result can be understood as follows. As argued in section \ref
{introduction}, $S_{I}(0)$ near threshold is dominated by the Poissonian
statistics of tunneling through a single bottleneck junction, and is thus
given by the Schottky formula: 
\begin{equation}
\left. \frac{S_{I}(0)}{2e\langle \bar{I}\rangle }\right| _{I\to 0}\to 1\,.
\end{equation}

\noindent At high voltages, however, a large number of charge states becomes
available for tunneling through each junction. Though the tunneling rate for
each of these states may be affected by the state of neighboring junctions,
these effects are averaged out, since the voltage dependence of the rate of
tunneling through each junction is linear at $\Delta W_{ij}>0$ (see Eqn. (%
\ref{gamma-zero})). Under these conditions, current noise through each
junction is described by the Schottky formula, $S_{I}^{(1)}(0)=2e\langle 
\bar{I}\rangle $ . Since we may transform current noise into voltage
noise by the square of the dynamic resistance (which at $V\to \infty $ just
equals $R$), we write the voltage noise of a single junction as 
\[
\left. S_{V}^{(1)}(0)\,\right| _{I\to \infty }=2e\langle \bar{I}\rangle
R^{2}\,.
\]

\noindent The total voltage noise $S_{V}$ is the simple sum of the noise of
the individual junctions $S_{V}^{(1)}$, while the total current noise is
finally obtained from $S_{V}$ via the total array resistance $NR$: 
\begin{eqnarray}
\left. S_{V}(0)\,\right| _{I\to \infty } &=&2e\langle \bar{I}\rangle R^{2}N,
\\
\left. S_{I}(0)\,\right| _{I\to \infty } &=&\frac{S_{V}(0)}{(NR)^{2}}=\frac{%
2e\langle \bar{I}\rangle }{N}\,.  \label{reduced}
\end{eqnarray}
(For the particular case $N=2$ this equation has been discussed in Ref. %
\onlinecite{jong}.) Thus the crossover in $S_{I}(0)/2e\langle \bar{I}\rangle 
$ from $1$ to $1/N$ with increasing current could be expected; what was
really surprising for us is that in most cases this crossover takes place
very close to the Coulomb blockade threshold, where the array $I-V$ curve is
still not quite linear, and hence the arguments given above cannot be taken
too seriously.

In order to describe the crossover quantitatively, we may define the {\it %
crossover current} $I_{c}$ as a value at which $S_{I}(0)/2e\langle \bar{I}%
\rangle $ is midway between these two limits, on a logarithmic scale: 
\begin{equation}
\frac{S_{I}(0)}{2eI_{c}}\equiv \frac{1}{\sqrt{N}}\,.
\end{equation}
We have written a Perl script called {\tt sicurve} to automate the
extraction of the crossover current $I_{c}$ from the $S-I$
curves. While invoking {\tt mso}, {\tt sicurve} continuously adjusts
the change in bias voltage between successive points, in an attempt to
produce a series of evenly spaced points in the
$S_{I}(0)/2e\langle\bar{I}\rangle$ vs. $\langle\bar{I}\rangle$ plane
(see Fig. \ref{s-i}). This is an important practical technique for
generating $S-I$ curves on circuits with random background charge,
since the relationship between voltage, current, and spectral density
can be quite irregular.

\section{Model 1: Arrays Near Ground Plane}

\label{ground-plane} Our first case was the simplest model\cite{bakh:1989}
of a uniform, symmetrically biased array near a ground plane with no
background charges (Fig. 1b). The direct capacitance matrix in this model is
tridiagonal, and is described by one dimensionless parameter, the ratio $%
C_{0}/C,$ where $C$ is the junction capacitance, and $C_{0}$ is the
``stray'' capacitance between an island and the ground plane. In this model,
the reciprocal length scale $\lambda =1/M$ of the single-electron soliton is
determined as 
\begin{equation}
\lambda =\cosh ^{-1}(C_{0}/2C-1)\,,
\end{equation}
and in the most interesting limit of $C_{0}\ll C$, $M=\sqrt{C/C_{0}}\gg 1$.

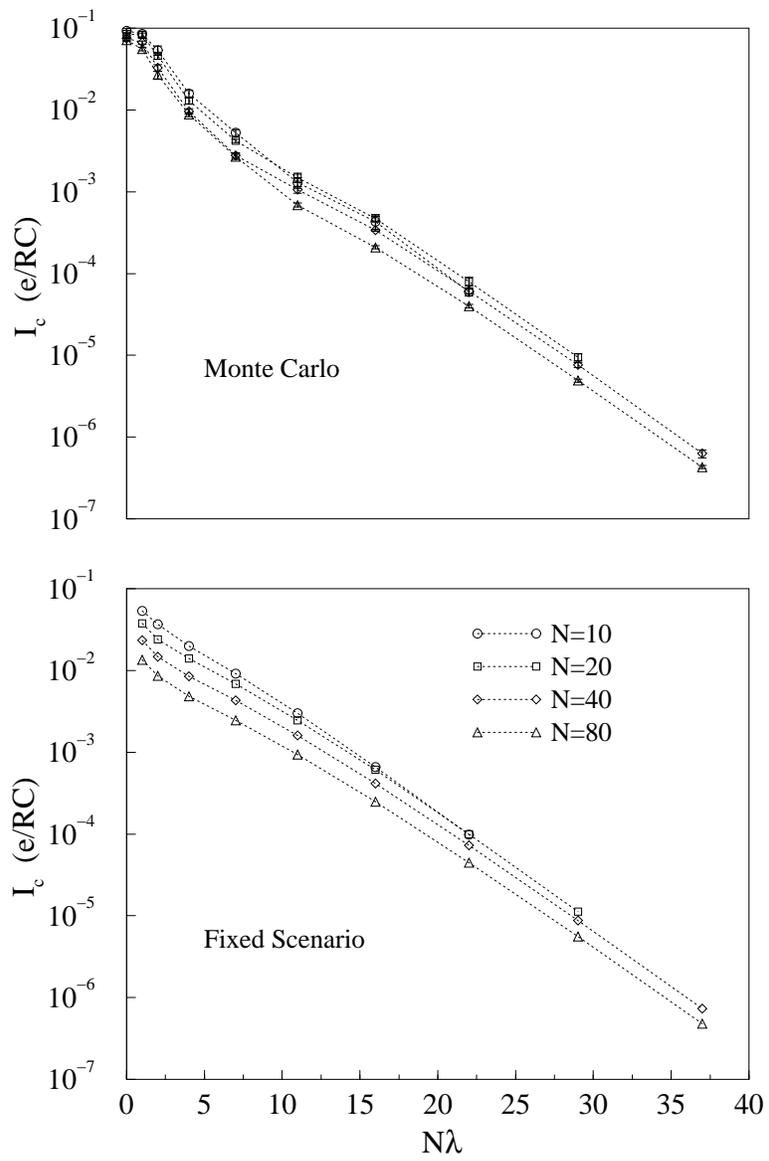
\begin{figure}
\caption{Crossover current $I_c$ as a function of the $N\lambda$, the ratio
of array length to charge soliton length, in arrays near ground plane with
no background charge.}
\label{tri-ic-nl}
\end{figure}

Surprisingly, our numerical results (Fig.~\ref{tri-ic-nl}) show that for all
values of $\lambda $ and $N,$ data for $I_{c}$ fall roughly on a single
universal curve, with \mbox{$I_c\propto\exp(-N\lambda/3)$} for %
\mbox{$\lambda\gtrsim 10$}. For a fixed product $N\lambda $, there is a
relatively weak decrease of $I_{c}$ with increasing $N$.

To understand this unexpected result, we began looking for an analytic
expression for current noise for the case when the passage of charge
through the array consists of a fixed, repeated sequence of tunneling
events. If such a sequence is repeated $n=Q/e$ times, we can write the
average total time as a sum of average times for each jump in the
fixed sequence:
\begin{equation}
\langle T_{Q}\rangle =n\sum_{i=1}^{N}\langle \delta t_{i}\rangle =\frac{Q}{e}%
\sum_{i=1}^{N}\Gamma _{i}^{-1}\,,
\end{equation}

\noindent where $\Gamma $ is the total rate at each stage of the process.
Since the time of each jump follows Poissonian statistics, 
\begin{equation}
{\rm Var}(\delta t_{i})=\Gamma _{i}^{-2}\,,
\end{equation}

\noindent and since the jump times are independent of one another, 
\begin{equation}
\mbox{\rm Var}(T_Q) = \frac{Q}{e}\sum_{i=1}^N \mbox{\rm Var}(\delta t_i) = 
\frac{Q}{e}\sum_{i=1}^N \Gamma_i^{-2}\,,
\end{equation}

\noindent we arrive at a simple formula\cite{noise}: 
\begin{equation}
\frac{S_{I}(0)}{2e\langle \bar{I}\rangle }=\frac{\displaystyle %
\sum_{i=1}^{N}\Gamma _{i}^{-2}}{\left( \displaystyle\sum_{i=1}^{N}\Gamma
_{i}^{-1}\right) ^{2}}\,\,.  \label{sian}
\end{equation}

This formula was obtained earlier\cite{noise} for a particular case $C_{0}=0$%
. In particular, it shows that the spectral density can be dominated by
bottleneck points, where the rate $\Gamma_{i}$ is much lower than average.

In order to find these bottleneck points for the present case of arrays near
a ground plane, let us examine the energy profile created by these arrays for
tunneling charges. The potential created by an electron in an array with
externals at both ends is\cite{bakh:1991} 
\begin{equation}
\phi _{s}(n,n^{\prime })=\frac{e}{C_{ef\!f}}\left\{ e^{-|n-n^{\prime }|\lambda
}-\frac{e^{-n\lambda }\sinh [(N-n^{\prime })\lambda ]+e^{-(N-n)\lambda
}\sinh (n\lambda )}{\sinh (N\lambda )}\right\} \,,  \label{tri-phi-s}
\end{equation}
where $n\in \{0,1\ldots N\}$ is the position of the electron in the array, $%
n^{\prime }$ is the measurement position, and $C_{ef\!f}\equiv \sqrt{%
C_{0}^{2}+4CC_{0}}$. For our symmetric bias ($V_{2}=-V_{1}=V/2$), the
potential created by the external electrodes is\cite{bakh:1991} 
\begin{equation}
\phi _{e}(n)=\frac{V}{2}\frac{\sinh [(N/2-n)\lambda ]}{\sinh (N\lambda /2)}%
\,.  \label{tri-v}
\end{equation}

Numerical simulations show that in symmetrically biased arrays with %
\mbox{$\lambda\gtrsim 1$}, the basic tunneling scenario near threshold is
the passage of electron-hole pairs. The components of the pair enter at
opposite ends of the array, move towards each other, then annihilate near
the center (see Table 1). We can write the energy of the electron-hole pair
as 
\begin{equation}
W(n_{1},n_{2})=W_{o}(n_{1})-e\phi _{e}(n_{1})+W_{o}(n_{2})+e\phi
_{e}(n_{2})-e\phi _{s}(n_{1},n_{2})\,,  \label{ehp-energy}
\end{equation}

\noindent where $n_{1}$ and $n_{2}$ are the positions of the electron and
hole, respectively, and 
\begin{equation}
W_{o}(n)=\frac{e\phi _{s}(n,n)}{2}=\frac{e}{2C_{ef\!f}}\left[ 1-\frac{\cosh
[(N-2n)\lambda ]-e^{-N\lambda }}{\sinh (N\lambda )}\right]
\end{equation}
is the self-energy of an electron or hole.

Figure \ref{tri-ic-nl} shows the results of calculation of the crossover
current using Eqs. (\ref{ehp-energy}) and (\ref{sian}) in a fixed scenario
picture, where the electron and hole enter the array one right after the
other, and then take turns tunneling towards each other in a symmetric
manner. For \mbox{$N\lambda\gtrsim 15$}, the results of this semi-analytical
calculation match the Monte Carlo results very closely. (For $N\lambda
\lesssim 15$, a difference appears, increasing with $N$.) This means that we
can analyze our problem, at least approximately, by examining the energy
profile in our fixed scenario. A straightforward analysis of Eq. (\ref
{ehp-energy}) shows that, for \mbox{$N\lambda\gg 1$}, there is a slow point
(a minimum in $\Delta W_{ij}$) when both electron and hole are %
\mbox{$\sim\!N/3$} islands from their respective edge of the array. Here
both members of the electron-hole pair are far from the edges and cannot be
pushed strongly by the external voltage, yet are not close enough to attract
each other strongly, either. At this point, $\Gamma $ scales as $\exp
(-N\lambda /3)$. According to Eq. (\ref{sian}), this leads to a similar
dependence of $I_{c}$, at least in the limit \mbox{$N\lambda\gg 1$}.

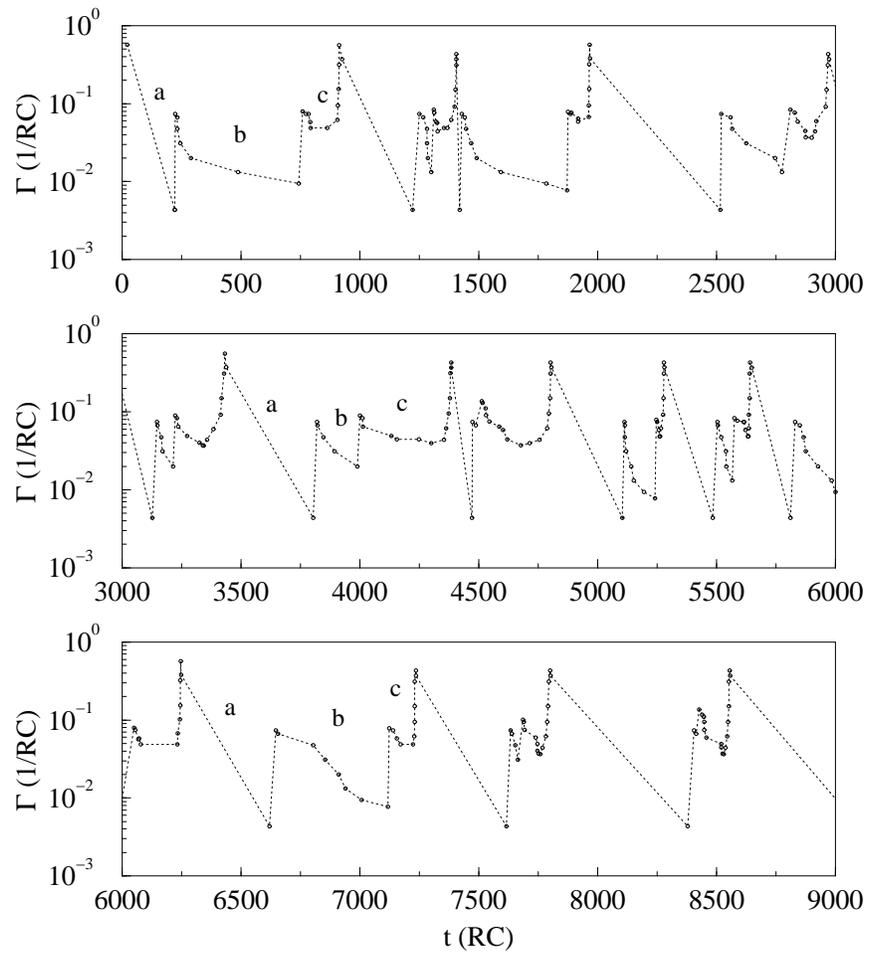
\begin{figure}
\caption{$\Gamma$ vs. time for a Monte Carlo simulation of an array near
ground plane with no background charge, symmetrically biased near threshold, 
$N=20, \lambda=0.55$. Slow points occur before the first charge enters the
array (a), and when either charge is $\sim N/3$ junctions away from the
nearest edge (b,c). Dotted line is only a guide for the eye.}
\label{tri-gt}
\end{figure}

Some difference between the fixed-scenario calculations and the Monte Carlo
simulations can be readily explained by the observation that frequently the
tunneling process is somewhat more complicated than the exactly alternating
electron and hole motion sequence---see Table 1. Figure~\ref{tri-gt} shows a
typical pattern of the total tunneling rates.

These figures show that sometimes there are three (rather then one) slow
points with low $\Gamma $. The first slow point is before any charge enters
the array (see labels (a) in Fig. \ref{tri-gt}). After the first charge
enters the array, the external voltage pushes it quickly away from the edge.
There is then typically a second slow point when that charge reaches a
distance of roughly $N/3$ islands from the external it emerged from (see
labels (b) in Fig. \ref{tri-gt}). Then the opposite charge enters the array
from the other side, also hopping quickly away from the edge. The third slow
point usually follows when the opposite charge is roughly $N/3$ islands away
from its external (see labels (c) in Fig. \ref{tri-gt}). Only the first two
points are taken into account in our fixed scenario. Since the second slow
point is frequently important, the good agreement between our Monte Carlo
and fixed scenario calculations is somewhat puzzling.

\section{Model 2: Arrays Without Ground Plane}

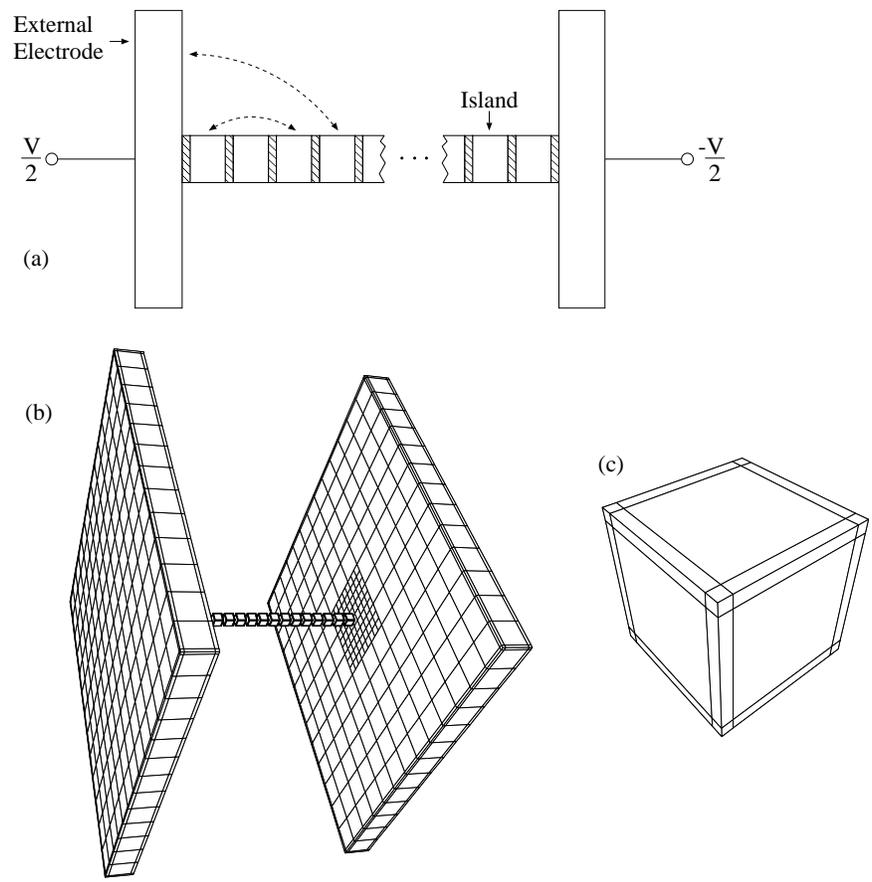
\begin{figure}
\caption{Array, without ground plane, between parallel plate externals. (a)
Schematics, (b) geometric model with surfaces divided into panels for
geometric capacitance calculation, and (c) closeup of a single island
paneling.}
\label{cube-diag}
\end{figure}

Here we examine the case of an array with islands between semi-infinite
external electrodes (Fig.~\ref{cube-diag}a), without a ground plane. In the
absence of a ground plane, mutual capacitances other than the junction
capacitances may become important.

To model the electrostatics of such arrays, we have used two
methods. In the first method, we create a geometric model of the
island and external electrodes (Fig. \ref{cube-diag}b), and use it to
calculate the full capacitance matrix for the array numerically using
FastCap\cite{fastcap}.  Since the electrostatics is rather insensitive
to the exact shape of the islands\cite{arrel}, they may be modeled by
cubes of side length $a$. The resulting capacitance matrix (an example
is given in Table ~\ref{cube-mat}) was used for the Monte Carlo
simulation of noise, as described above.

In the second method, we used a simple heuristic approximation for the
single-electron soliton potential at distance $m=|n-n^{\prime}|$ in
a long array, found in Ref. \onlinecite{arrel}, 
\begin{equation}
\phi _{s}(m)=\frac{e}{a}\left\{ \alpha \lambda \exp (-\kappa \lambda m)+%
\frac{1}{m}\left[ 1-\exp (-\kappa \lambda m)\right] \right\} \,,
\label{simpexp}
\end{equation}
with $\alpha \approx \kappa \approx 1$. This formula describes a crossover
from an exponential decay at short distance ($\lambda m<1)$ to a Coulomb-law 
$1/r$ decay at large distance. The effect of the external electrodes was
described by the usual image charge method; in our case, with two
electrodes, it involves an infinite series of images. As a result, the full
single electron potential, $\phi _{s}^{h}(n,n^{\prime })$, may be expressed
as the sum of an infinite series over all image charge contributions, and
the self energy can be written as 
\begin{eqnarray*}
W_{o}^{h}(n) &=&\frac{e\phi _{s}^{h}(n,n)}{2} \\
&=&\sum_{i=1}^{\infty }\left\{ W_{a}(2Ni)-\frac{%
[W_{a}(2N(i-1)+2n)+W_{a}(2Ni-2n)]}{2}\right\} \,.
\end{eqnarray*}
On the other hand, the form of the external potential is quite simple, 
\[
\phi _{e}^{h}(n)\simeq eV(N-2n)/2N\,.
\]
The energy for an electron-hole pair can then be written just as in Eqn. (%
\ref{ehp-energy}). This energy was used for the fixed-scenario calculation
of noise, similar to that described in the previous section.

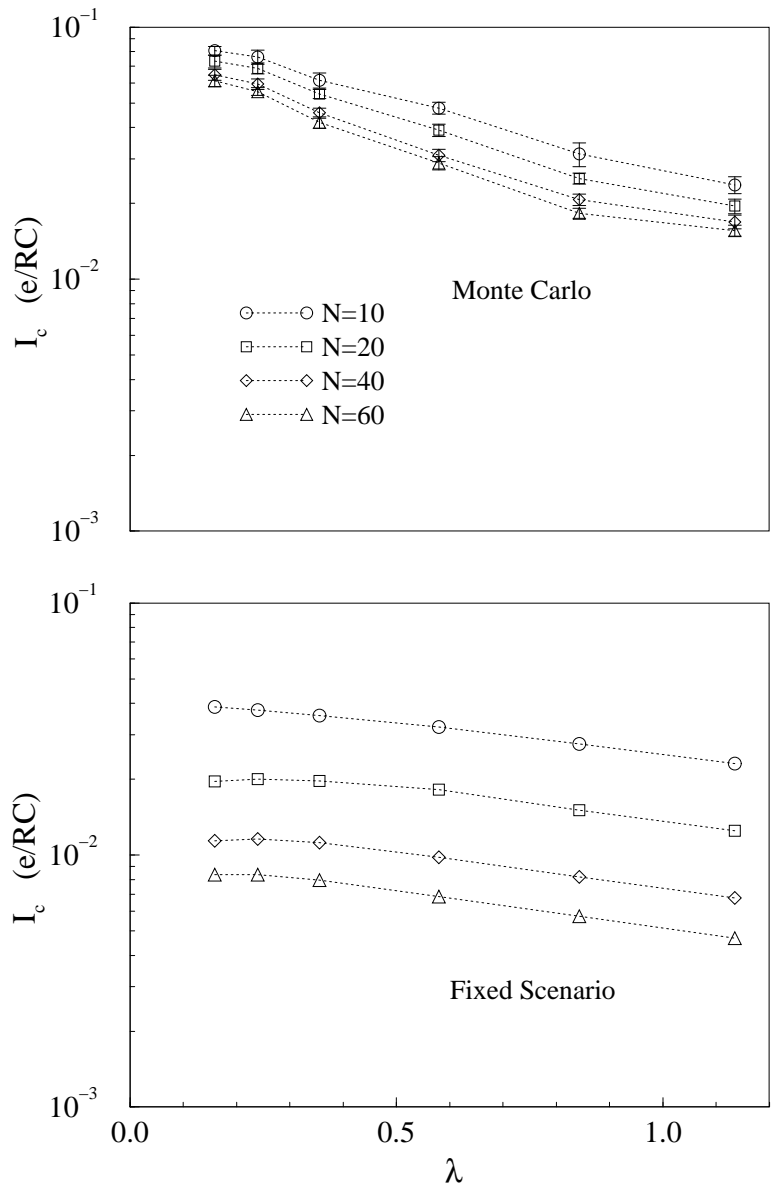
\begin{figure}
\caption{Crossover current vs. inverse soliton length for arrays without
ground plane between parallel plate externals.}
\label{cube-ic-l}
\end{figure}

Results from both methods are shown in Fig.~\ref{cube-ic-l}. The Monte Carlo
simulation shows that, unlike in arrays near a ground plane, $I_{c}$ does
not follow a universal dependence on $N\lambda $. Rather, the $I_{c}$ values
follow a common curve depending mostly on $\lambda $ alone, decreasing
weakly with the array length $N$. Even the $\lambda $ dependence is weak,
compared to the results for arrays near ground plane: for $N=20$ in the
range of $\lambda $ from 0.25 to 1, $I_{c}$ drops by a factor of $\sim \!3$
in arrays with no ground plane, whereas it drops by more than two decades in
arrays with ground plane (Fig.~\ref{tri-ic-nl}).

The fixed scenario results for $I_{c}$ match the Monte Carlo results
fairly well in terms of the shape of the $\lambda $ dependence of the
curves. For $N=10$, the magnitude of the results are also in fairly
close agreement. However, the fixed scenario results exhibit a
stronger decline with $N$ than the Monte Carlo results. For $N=60$,
the fixed scenario values of $I_{c}$ are between 3 and 8 times smaller
than the corresponding Monte Carlo values.

Generally, it is easy to understand why the fixed scenario values for $I_{c}$
fall below the values calculated with Monte Carlo simulation: randomness of
jump location, which is ignored in the fixed scenario calculation, can be
thought of as an additional source of noise. As the general level of noise
in the crossover region increases, so does the crossover current. However,
we are still in need of a simple interpretation of the $N$ and $\lambda $
dependence of the shot noise in arrays without ground plane.

\section{Model 3: Arrays with Random Background Charge}

\label{q0-arrays} Returning to arrays near ground plane, with their simple
electrostatics, we have explored the behavior of shot noise in the
presence of random background charges on the islands,
$\vec{q}_{0}$. These charges can represent, for example, the effect of
charged impurities in a substrate.  The charge $q_{0,i}$ placed on
each island was randomly selected\cite {recipes}, using a uniform
probability distribution between $-e/2$ and $e/2$ . (Any integer part
of the background charge would immediately be compensated by trapping
one or a few tunneling electrons or holes.)

The $S-I$ curves which stem from the Monte Carlo simulation of such arrays
still show the progression from shot noise to suppressed shot noise, but at
larger values of $\lambda $ tend to feature strong, irregular peaks (Fig. 
\ref{s-i}) where curves for arrays without background charge were smooth.
Occasionally these peaks in the $S-I$ curve cause it to cross the $1/\sqrt{N}
$ line more than once. In such cases, we arbitrarily define $I_{c}$ as the
lowest crossing value. It turns out that the variation in the general
location of $I_{c}$ among different instantiations of $\vec{q}_{0}$ is large
enough to make the distinctions among different crossing points on a single
curve irrelevant.

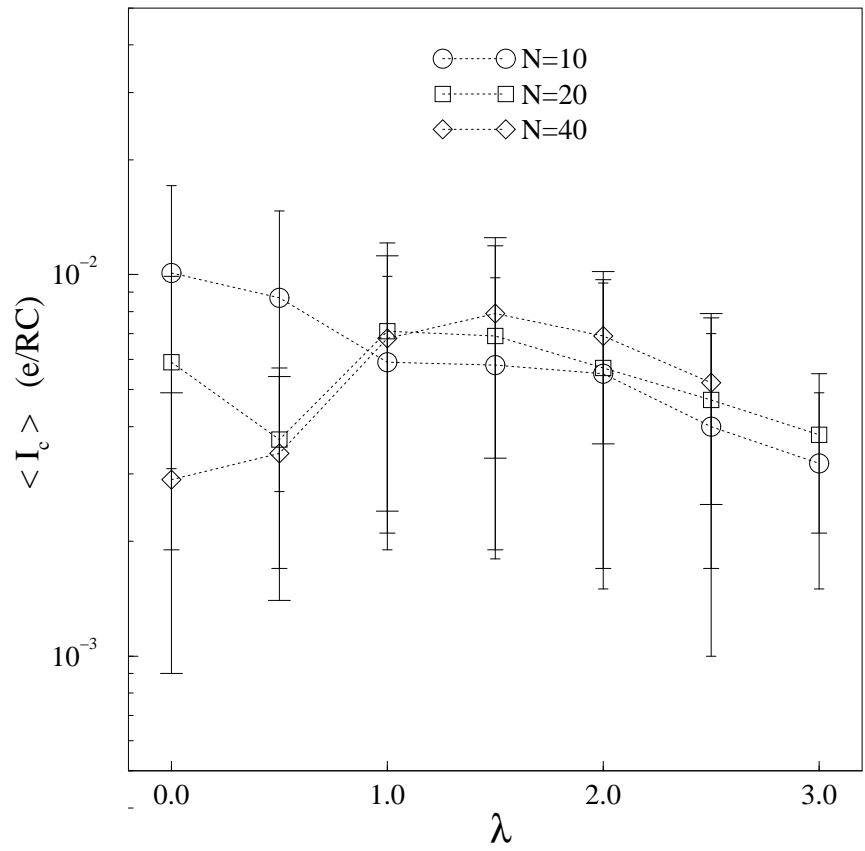
\begin{figure}
\caption{Crossover current as a function of inverse soliton length in arrays
near ground plane with random background charge. Each point with its error
bars represents the average and standard deviation over ensembles of fifty
different random background charge distributions. Error bar widths are
scaled for readability: narrowest for $N=10$, widest for $N=40$.}
\label{q0-ic-l}
\end{figure}

Figure \ref{q0-ic-l} shows the average value of $I_{c}$ and its standard
deviation of over ensembles of 50 different $\vec{q}_{0}$ realizations.
Strikingly, for $\lambda \gtrsim 1$, $I_{c}$ for all values of $N$ appears
to fall on the same, almost flat curve. This behavior is in sharp contrast
to results for similar arrays without background charge (Fig.~\ref{tri-ic-nl}%
). This quasi-universal value of $I_{c}$ is $\sim \!5\times 10^{-3}e/RC$.
Although there appears to be weak downward trend in $\langle I_{c}\rangle $
with increasing $\lambda $ for \mbox{$\lambda\gtrsim 1.5$}, this trend is
virtually hidden within the relatively large standard deviation.

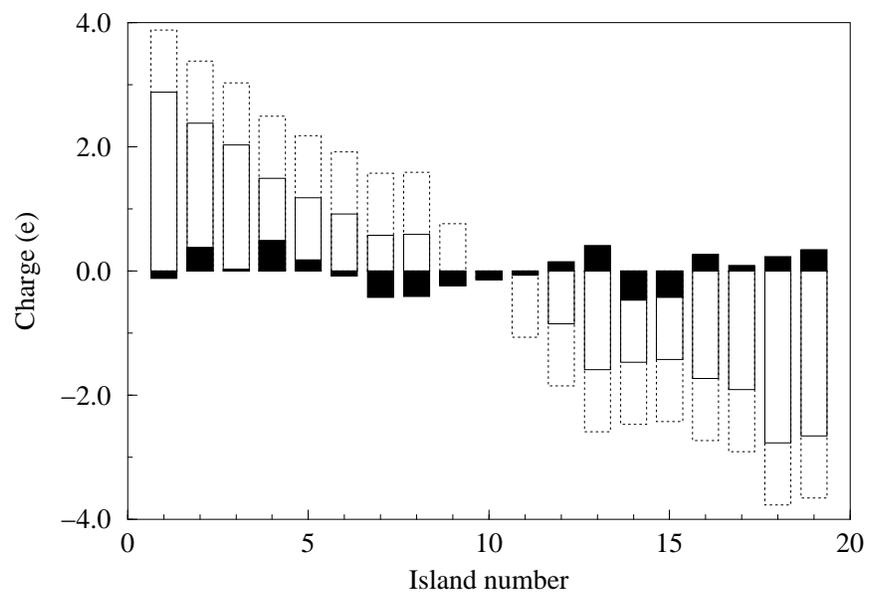
\begin{figure}
\caption{Typical charge distribution in array with random background charges 
$\vec{q}_{0}$. \mbox{$N=20$}, \mbox{$\lambda=2$}. Filled bars represent $%
\vec{q}_{0}$, solid lines represent the sandpile at \mbox{$V=V_{t}$}, and
dotted lines represent conduction charges in a simple alternating 
\mbox{$e-h$} scenario.}
\label{sandpile}
\end{figure}

Let us try to comprehend this result. Simulation shows that in such
arrays the typical tunneling process near the crossover is due to
several (rather than one) electron-hole pairs moving simultaneously in
the array. (Due to this multiplicity, a fixed-scenario calculation of
noise would not make sense.) This effect is easy to explain. The
electric potential induced by the background charges creates a series
of charge traps. So, as the applied voltage is increased, the first
several charges to enter the array are trapped, forming a
``sandpile''\cite{middleton} of charge (see Fig.~\ref{sandpile} ). The
final energy profile in the sandpile is still random, though the
maximum possible energy change ${\Delta W}_{{ij}}$ is now upwardly
bound by $\Delta W_{max}\approx e^{2}/2C_{ef\!f}$\,. The remaining
disorder is, however, strong enough to overcome the interaction of
distant charges, which is exponentially weak in arrays over a ground
plane.

This argument also explains why the crossover current is independent of the
array length. We are, however, still in need of an analytical theory which
would explain the virtual independence of the average crossover on $\lambda$%
\,, and also would predict its universal value cited above.

\section{Conclusion}

We have used both Monte Carlo and fixed scenario techniques to
calculate the low-frequency current noise for three different models
of 1D single-electron-tunneling arrays. Within each of the three
models, we find a crossover of the spectral density of current
fluctuations, $S_{I}(0)$, from the Schottky value $2e\langle
\bar{I}\rangle $, to the ``reduced Schottky'' value $2e\langle
\bar{I}\rangle /N$, with increasing current. The crossover can be well
characterized in terms of the crossover current $I_{c}$, which may be
said to mark the onset of quasicontinuous charge transfer. The
particular behavior of $I_{c}$ as a function of $N$ and $\lambda $
depends on the interaction of electrons within the array and on the
interaction of electrons with the external field.

For arrays near ground plane, with no background charge, the crossover
current exhibits a universal behavior that is a function of only the product 
$N\lambda $\thinspace , i.e. of the ratio of the array length $N$ to the
length $M=\lambda ^{-1}$ of single-electron solitons. At $N\lambda >10,$ the
dependence is exponential: $I_{c}\propto \exp (-N\lambda /3)$. Our analysis
has shown that this behavior is the result of the exponential decrease of
the soliton interaction with the external electrodes and its counterpart in
the electron-hole pair.

In arrays without a ground plane, $I_{c}$ is almost independent of $N$
and exhibits a nearly-universal weak decrease with $\lambda$.  We
believe that the substantial difference in results between this model
and the previous model is due to the long-range electrostatic
interactions, which were screened by the ground plane in the previous
model.

Finally, in arrays with random $\vec{q}_{0}$, we have found that the
crossover takes place at a nearly universal value of current,
$I_{c}\sim 5\times 10^{-3}e/RC$. The absence of noise dependence on
the array length can be readily explained as a result of the random
potential created by the background charges, which overwhelms long
range order in the arrays.  However, the independence of noise on the
single-electron soliton length still has to be explained.

\section{Acknowledgments}

The authors thank D.V. Averin and A.N. Korotkov for fruitful discussions,
and A. Huq for technical assistance.

\newpage

\begin{table}[tbp]
\begin{center}
{
\scriptsize
\renewcommand{\arraystretch}{0.7}
$\begin{array}{lrrrrrrrrrrrrrrrrrrrrrrrrrrrrrrrrrrrrrrr}
1.574463e+07 & . & . & . & . & . & . & . & . & . & . & . & . & . & . & . & .
& . & . & . & . & . & . & . & . & . & . & . & . & . & . & . & . & . & . & .
& . & . & . & . \\ 
1.832335e+07 & e & . & . & . & . & . & . & . & . & . & . & . & . & . & . & .
& . & . & . & . & . & . & . & . & . & . & . & . & . & . & . & . & . & . & .
& . & . & . & . \\ 
1.832340e+07 & . & e & . & . & . & . & . & . & . & . & . & . & . & . & . & .
& . & . & . & . & . & . & . & . & . & . & . & . & . & . & . & . & . & . & .
& . & . & . & . \\ 
1.832342e+07 & . & . & e & . & . & . & . & . & . & . & . & . & . & . & . & .
& . & . & . & . & . & . & . & . & . & . & . & . & . & . & . & . & . & . & .
& . & . & . & . \\ 
1.832412e+07 & . & . & . & e & . & . & . & . & . & . & . & . & . & . & . & .
& . & . & . & . & . & . & . & . & . & . & . & . & . & . & . & . & . & . & .
& . & . & . & . \\ 
1.832451e+07 & . & . & . & . & e & . & . & . & . & . & . & . & . & . & . & .
& . & . & . & . & . & . & . & . & . & . & . & . & . & . & . & . & . & . & .
& . & . & . & . \\ 
1.832525e+07 & . & . & . & . & . & e & . & . & . & . & . & . & . & . & . & .
& . & . & . & . & . & . & . & . & . & . & . & . & . & . & . & . & . & . & .
& . & . & . & . \\ 
1.832588e+07 & . & . & . & . & . & . & e & . & . & . & . & . & . & . & . & .
& . & . & . & . & . & . & . & . & . & . & . & . & . & . & . & . & . & . & .
& . & . & . & . \\ 
1.832873e+07 & . & . & . & . & . & . & . & e & . & . & . & . & . & . & . & .
& . & . & . & . & . & . & . & . & . & . & . & . & . & . & . & . & . & . & .
& . & . & . & . \\ 
1.833854e+07 & . & . & . & . & . & . & . & . & e & . & . & . & . & . & . & .
& . & . & . & . & . & . & . & . & . & . & . & . & . & . & . & . & . & . & .
& . & . & . & . \\ 
1.834973e+07 & . & . & . & . & . & . & . & . & . & e & . & . & . & . & . & .
& . & . & . & . & . & . & . & . & . & . & . & . & . & . & . & . & . & . & .
& . & . & . & . \\ 
1.844754e+07 & . & . & . & . & . & . & . & . & . & . & e & . & . & . & . & .
& . & . & . & . & . & . & . & . & . & . & . & . & . & . & . & . & . & . & .
& . & . & . & . \\ 
1.861754e+07 & . & . & . & . & . & . & . & . & . & . & . & e & . & . & . & .
& . & . & . & . & . & . & . & . & . & . & . & . & . & . & . & . & . & . & .
& . & . & . & . \\ 
1.899889e+07 & . & . & . & . & . & . & . & . & . & . & . & e & . & . & . & .
& . & . & . & . & . & . & . & . & . & . & . & . & . & . & . & . & . & . & .
& . & . & . & h \\ 
1.899894e+07 & . & . & . & . & . & . & . & . & . & . & . & e & . & . & . & .
& . & . & . & . & . & . & . & . & . & . & . & . & . & . & . & . & . & . & .
& . & . & h & . \\ 
1.899896e+07 & . & . & . & . & . & . & . & . & . & . & . & e & . & . & . & .
& . & . & . & . & . & . & . & . & . & . & . & . & . & . & . & . & . & . & .
& . & h & . & . \\ 
1.899896e+07 & . & . & . & . & . & . & . & . & . & . & . & e & . & . & . & .
& . & . & . & . & . & . & . & . & . & . & . & . & . & . & . & . & . & . & .
& h & . & . & . \\ 
1.899899e+07 & . & . & . & . & . & . & . & . & . & . & . & e & . & . & . & .
& . & . & . & . & . & . & . & . & . & . & . & . & . & . & . & . & . & . & h
& . & . & . & . \\ 
1.899899e+07 & . & . & . & . & . & . & . & . & . & . & . & e & . & . & . & .
& . & . & . & . & . & . & . & . & . & . & . & . & . & . & . & . & . & h & .
& . & . & . & . \\ 
1.900215e+07 & . & . & . & . & . & . & . & . & . & . & . & e & . & . & . & .
& . & . & . & . & . & . & . & . & . & . & . & . & . & . & . & . & h & . & .
& . & . & . & . \\ 
1.900377e+07 & . & . & . & . & . & . & . & . & . & . & . & e & . & . & . & .
& . & . & . & . & . & . & . & . & . & . & . & . & . & . & . & h & . & . & .
& . & . & . & . \\ 
1.902246e+07 & . & . & . & . & . & . & . & . & . & . & . & e & . & . & . & .
& . & . & . & . & . & . & . & . & . & . & . & . & . & . & h & . & . & . & .
& . & . & . & . \\ 
1.906005e+07 & . & . & . & . & . & . & . & . & . & . & . & e & . & . & . & .
& . & . & . & . & . & . & . & . & . & . & . & . & . & h & . & . & . & . & .
& . & . & . & . \\ 
1.908176e+07 & . & . & . & . & . & . & . & . & . & . & . & e & . & . & . & .
& . & . & . & . & . & . & . & . & . & . & . & . & h & . & . & . & . & . & .
& . & . & . & . \\ 
1.918411e+07 & . & . & . & . & . & . & . & . & . & . & . & e & . & . & . & .
& . & . & . & . & . & . & . & . & . & . & . & h & . & . & . & . & . & . & .
& . & . & . & . \\ 
1.926408e+07 & . & . & . & . & . & . & . & . & . & . & . & . & e & . & . & .
& . & . & . & . & . & . & . & . & . & . & . & h & . & . & . & . & . & . & .
& . & . & . & . \\ 
1.995202e+07 & . & . & . & . & . & . & . & . & . & . & . & . & . & e & . & .
& . & . & . & . & . & . & . & . & . & . & . & h & . & . & . & . & . & . & .
& . & . & . & . \\ 
2.001400e+07 & . & . & . & . & . & . & . & . & . & . & . & . & . & e & . & .
& . & . & . & . & . & . & . & . & . & . & h & . & . & . & . & . & . & . & .
& . & . & . & . \\ 
2.005999e+07 & . & . & . & . & . & . & . & . & . & . & . & . & . & . & e & .
& . & . & . & . & . & . & . & . & . & . & h & . & . & . & . & . & . & . & .
& . & . & . & . \\ 
2.024492e+07 & . & . & . & . & . & . & . & . & . & . & . & . & . & . & e & .
& . & . & . & . & . & . & . & . & . & h & . & . & . & . & . & . & . & . & .
& . & . & . & . \\ 
2.025450e+07 & . & . & . & . & . & . & . & . & . & . & . & . & . & . & . & e
& . & . & . & . & . & . & . & . & . & h & . & . & . & . & . & . & . & . & .
& . & . & . & . \\ 
2.025463e+07 & . & . & . & . & . & . & . & . & . & . & . & . & . & . & . & e
& . & . & . & . & . & . & . & . & h & . & . & . & . & . & . & . & . & . & .
& . & . & . & . \\ 
2.025538e+07 & . & . & . & . & . & . & . & . & . & . & . & . & . & . & . & .
& e & . & . & . & . & . & . & . & h & . & . & . & . & . & . & . & . & . & .
& . & . & . & . \\ 
2.025579e+07 & . & . & . & . & . & . & . & . & . & . & . & . & . & . & . & .
& e & . & . & . & . & . & . & h & . & . & . & . & . & . & . & . & . & . & .
& . & . & . & . \\ 
2.025601e+07 & . & . & . & . & . & . & . & . & . & . & . & . & . & . & . & .
& e & . & . & . & . & . & h & . & . & . & . & . & . & . & . & . & . & . & .
& . & . & . & . \\ 
2.025622e+07 & . & . & . & . & . & . & . & . & . & . & . & . & . & . & . & .
& e & . & . & . & . & h & . & . & . & . & . & . & . & . & . & . & . & . & .
& . & . & . & . \\ 
2.025626e+07 & . & . & . & . & . & . & . & . & . & . & . & . & . & . & . & .
& . & e & . & . & . & h & . & . & . & . & . & . & . & . & . & . & . & . & .
& . & . & . & . \\ 
2.025627e+07 & . & . & . & . & . & . & . & . & . & . & . & . & . & . & . & .
& . & . & e & . & . & h & . & . & . & . & . & . & . & . & . & . & . & . & .
& . & . & . & . \\ 
2.025629e+07 & . & . & . & . & . & . & . & . & . & . & . & . & . & . & . & .
& . & . & . & e & . & h & . & . & . & . & . & . & . & . & . & . & . & . & .
& . & . & . & . \\ 
2.025630e+07 & . & . & . & . & . & . & . & . & . & . & . & . & . & . & . & .
& . & . & . & e & h & . & . & . & . & . & . & . & . & . & . & . & . & . & .
& . & . & . & . \\ 
2.025630e+07 & . & . & . & . & . & . & . & . & . & . & . & . & . & . & . & .
& . & . & . & . & . & . & . & . & . & . & . & . & . & . & . & . & . & . & .
& . & . & . & .
\end{array}$
}
\end{center}
\caption{Conduction diagram in a Monte Carlo simulation of an array near a
ground plane with no background charge, symmetrically biased near threshold, 
$N=40,\lambda =0.96$. The first column is time, subsequent columns show the
charge on each island.}
\label{tri-cond}
\end{table}

\begin{table}[tbp]
\begin{center}
{\scriptsize
\renewcommand{\arraystretch}{0.8}
$\begin{array}{rrrrrrrrrrr}
& 3~~~~ & 4~~~~ & 5~~~~ & 6~~~~ & 7~~~~ & 8~~~~ & 9~~~~ & 10~~~~ & 11~~~~ & 
12~~~~ \\ 
1 & 2.99525 & 0.33124 & 0.25056 & 0.20843 & 0.18008 & 0.15821 & 0.13997 & 
0.12453 & 0.11055 & 0.09774 \\ 
2 & 0.00456 & 0.01429 & 0.02383 & 0.03347 & 0.04319 & 0.05314 & 0.06333 & 
0.07441 & 0.08573 & 0.09768 \\ \hline
3 &  &  &  &  &  &  &  &  &  &  \\ 
4 & 2.56553 &  &  &  &  &  &  &  &  &  \\ 
5 & -0.01817 & 2.59339 &  &  &  &  &  &  &  &  \\ 
6 & -0.00665 & -0.02876 & 2.59838 &  &  &  &  &  &  &  \\ 
7 & -0.00319 & -0.01199 & -0.03172 & 2.60029 &  &  &  &  &  &  \\ 
8 & -0.00181 & -0.00632 & -0.01391 & -0.03303 & 2.60122 &  &  &  &  &  \\ 
9 & -0.00115 & -0.00385 & -0.00764 & -0.01485 & -0.03372 & 2.60174 &  &  & 
&  \\ 
10 & -0.00070 & -0.00267 & -0.00473 & -0.00834 & -0.01537 & -0.03412 & 
2.60205 &  &  &  \\ 
11 & -0.00050 & -0.00172 & -0.00323 & -0.00531 & -0.00874 & -0.01567 & 
-0.03434 & 2.60223 &  &  \\ 
12 & -0.00037 & -0.00122 & -0.00228 & -0.00357 & -0.00562 & -0.00897 & 
-0.01584 & -0.03448 & 2.60231 &  \\ 
13 & -0.00029 & -0.00097 & -0.00163 & -0.00259 & -0.00387 & -0.00576 & 
-0.00902 & -0.01595 & -0.03452 & 2.60231 \\ 
14 & -0.00022 & -0.00077 & -0.00130 & -0.00186 & -0.00270 & -0.00405 & 
-0.00611 & -0.00906 & -0.01591 & -0.03447 \\ 
15 & -0.00017 & -0.00055 & -0.00096 & -0.00142 & -0.00200 & -0.00281 & 
-0.00407 & -0.00585 & -0.00910 & -0.01585 \\ 
16 & -0.00013 & -0.00041 & -0.00073 & -0.00107 & -0.00149 & -0.00203 & 
-0.00276 & -0.00400 & -0.00580 & -0.00897 \\ 
17 & -0.00010 & -0.00032 & -0.00055 & -0.00080 & -0.00110 & -0.00149 & 
-0.00197 & -0.00276 & -0.00382 & -0.00562 \\ 
18 & -0.00009 & -0.00026 & -0.00041 & -0.00057 & -0.00076 & -0.00100 & 
-0.00137 & -0.00190 & -0.00265 & -0.00365 \\ 
19 & -0.00002 & -0.00011 & -0.00025 & -0.00042 & -0.00063 & -0.00088 & 
-0.00121 & -0.00121 & -0.00154 & -0.00223 \\ 
20 & -0.00002 & -0.00008 & -0.00015 & -0.00023 & -0.00033 & -0.00046 & 
-0.00063 & -0.00069 & -0.00092 & -0.00124 \\ 
21 & -0.00001 & -0.00003 & -0.00005 & -0.00007 & -0.00010 & -0.00014 & 
-0.00019 & -0.00021 & -0.00028 & -0.00037
\end{array}$
}
\end{center}
\caption{Partial capacitance matrices generated by FastCap for a cubic array
with $N=20$ cubes between large parallel plates. Rows 1 and 2 belong to {%
\protect\boldmath $\tilde{C}$}, rows 3-21 belong to {\protect\boldmath $C$}.
By convention, all $\tilde{C}_{ij}>0$. All values in $10^{-16}$F. The array
parameters are $b=1.0$, $d=0.04$, $\lambda =0.356$.}
\label{cube-mat}
\end{table}

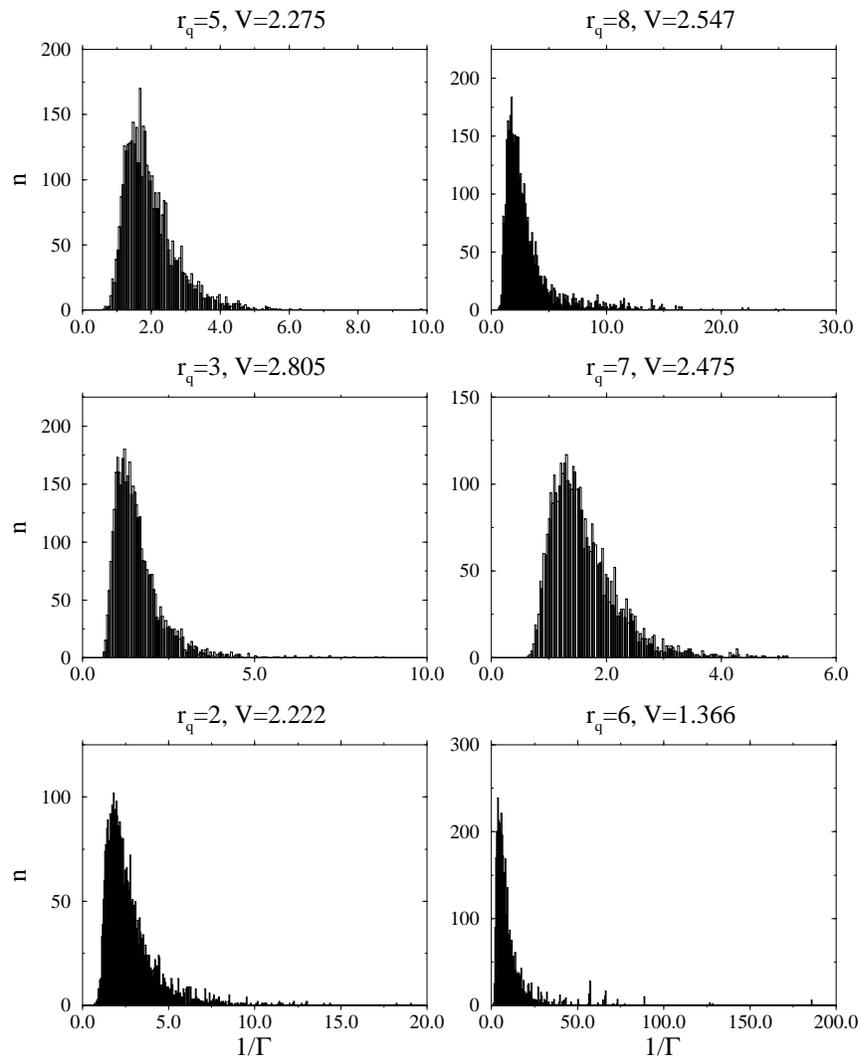
\begin{figure}
\caption{Histograms of the inverse total tunneling rate $1/\Gamma $ at $%
I=I_{c}$ for an array near ground plane ($N=40,\lambda =0.995$) with several
different random charge distributions. $r_{q}$ is the seed given to the
random number generator for background charges.}
\label{gamma-hist}
\end{figure}

\end{document}